\documentstyle[preprint,tighten,aps]{revtex} 
\begin{document}
\draft
\preprint{\vbox{\noindent
\null\hfill INFNCA-TH9710\\
           \null\hfill astro-ph/9710173}}
\title{Anomalous diffusion modifies solar neutrino fluxes}
\author{G. Kaniadakis$^{1,2,}$\cite{email1},
        A. Lavagno$^{1,2,}$\cite{email2}, 
        M. Lissia$^{3,4,}$\cite{email3}
    and P. Quarati$^{1,3,}$\cite{email4}}
\address{
$^{1}$Dipartimento di Fisica and INFM, Politecnico di Torino,
      I-10129 Torino, Italy \\
$^{2}$Istituto Nazionale di Fisica Nucleare, Sezione di Torino\\
$^{3}$Istituto Nazionale di Fisica Nucleare, Sezione di Cagliari,\\
      C.P. 170, I-09042 Monserrato, Italy\\
$^{4}$Dipartimento di Fisica, Universit\`a di Cagliari,
                I-09042 Monserrato, Italy\\
        }
\date{8 October 1997; revised version 24 April 1998}
\maketitle
\begin {abstract} 
    Density and temperature conditions in the solar core suggest that
 the microscopic diffusion of electrons and ions could be nonstandard:
 Diffusion and friction coefficients are energy dependent,
 collisions are not two-body processes and retain memory beyond
 the single scattering event.
    A direct consequence of nonstandard diffusion is that the equilibrium
 energy distribution of particles departs from the Maxwellian one
    (tails goes to zero more slowly or faster than exponentially)
modifying the reaction rates.
 This effect is qualitatively different from
 temperature and/or composition modification: Small changes in the
 number of particles in the distribution tails can strongly modify
 the rates without affecting bulk properties, such as the sound speed
 or hydrostatic equilibrium, which depend on the mean values from the
 distribution.
    This mechanism can considerably increase the range of predictions for
 the neutrino fluxes allowed by the current experimental values (cross
 sections and solar properties) and can be used to reduce the discrepancy
 between these predictions and the solar neutrino experiments.
\end{abstract}

\pacs{26.65.+t, 05.20.-y, 65.50.+m}
\narrowtext
\section{Introduction}
\label{sec:intr}
 Nowadays solar modeling seems to have reached a satisfactory
   stage~\cite{clayton68,bahcall89,bahcall95a,bahcall95b}.
   The inclusion in the latest models of higher-order effects, such as the
   diffusion of heavy elements, brings the theoretical predictions in good
   agreement even with the detailed helioseismological
   data~\cite{bahcall97b,fiorentini97}.

 Nevertheless, solar neutrino experiments still contradict the model
  predictions. The problem appears to be almost independent of the
  details of solar
   models~\cite{bahcall90,white93,castell93,hata94a,castell94b,%
bahcall94,parke95,bere96,castell97}, 
   and it has mostly been interpreted as a hint of new
physics~\cite{bahcall95b,fukugita91,giunti92,GALLEX92,Bludman93,hata93,%
Krastev93,bilenkii94,fiore94,bilenkii95,Krastev96a,Krastev96b,%
Krastev97,castell94b}.
   However, not everybody believes that the discrepancy is large enough
   (3$\sigma$ effects have often disappeared)
   and that the solar models, in spite of their successes, are still solid
   enough to pose a real problem~\cite{dar97}.
   
  In this context, there has been a considerable amount of work devoted
   to answering questions such as: how large are the uncertainties of the
   solar model input parameters? Has something been left out of standard
   solar models? How does this affects predictions for the fluxes and the
   status of solar neutrino 
   problem (SNP)~\cite{Krauss93,Fogli94,hata94b,fogli95,bahcall96,castell97}?

 The solar core is a dense, strongly interacting many-body system: no
   realistic microscopic calculation of such a system exists. However,
   only three properties are important for the purpose of solar modeling.
   The first one is the equation of state, {\em i.e.}, the relation 
   between the average one-body local properties of the system
   (temperature, pressure, density). It is the equation of state that
   regulates the local hydrodynamical equilibrium and, in addition,
   contributes to the interpretation of the helioseismological data.
   The second one is the local opacity, which controls the energy
   transmission.
   In this paper we are interested in the last feature, the two-body
   relative-energy distribution or two-body correlation between particles.
   The rates of the most important reactions in the Sun are strongly affected
   by the high-energy tail of this distribution.

  Solar models implicitly assume that the solar core can be described in
   terms of a gas of particles interacting via two-body short-range forces
   with no many-body effects apart for mean-field screening. Therefore,
   the velocity distribution and, in particular, the relative velocity
   distribution for the particles involved in the reactions are
   Maxwellian.

   Two decades ago, Kocharov and
   collaborators~\cite{kocha72,kocha74,kocha75},
   and Clayton and collaborators~\cite{clayton74,clayton75}
   speculated that the high-energy tail of the relative energy
   distribution could depart from the Maxwell-Boltzmann (MB) exponential
   form, $\exp\{ - E / kT \}$.
   At that time the SNP consisted only in the low yield of the Chlorine
   experiment~\cite{davis73} compared to theoretical
   predictions~\cite{bahcall69}. Since only the highest-energy proton
   can significantly penetrate the Coulomb barrier of the reaction
  $^7$Be$(p,\gamma){}^8$B, Clayton suggested that a significantly lower
   boron-neutrino
   flux could be obtained by depleting the small number of protons in the
   tail and proposed to parameterize the small deviation with a Gaussian
   factor, $\exp\{ -\delta (E/kT)^2\}$: a value of $\delta=0.01$ was
   sufficient to solve the SNP. This suggestion met some criticism and
   has been mostly ignored (see, however, a few recent
   papers~\cite{haubold95a,haubold95b,kania96,quarati97}). In particular,
   it has been remarked that deviations
   from the MB energy distribution could not be due to nuclear
   reactions keeping the distribution out of 
   equilibrium~\cite{rosenbluth57,macdonald57,krook76,bahcall89}.
   In addition, nobody was able to give a microscopical dynamical
   derivation of a non-Maxwellian distribution for the specific conditions
   inside the Sun (however, nobody has demonstrated with realistic
   calculations, which do not already assume the validity of the Boltzmann
   transport equation, that the distribution in the Sun is Maxwellian either).

 Several new developments have convinced us that it is necessary to
   reconsider the possibility of deviation from a MB distribution.
   First of all, non-Maxwellian {\em equilibrium} energy distributions
   have been shown theoretically
   possible~\cite{tsallis88,curado91,plastino94,stariolo94,tsallis95a}
   and relevant to many physical
   systems~\cite{plastino93,tsallis95b,zanette95,tsallis95c,%
tsallis96b,bogho96,hamity96,rajagopal96,costa97,torres97,lavagno97}.
   Then, it has been recently proved by explicitly solving the
   Fokker-Planck equation that a velocity-dependent diffusion or friction
   coefficient results in an equilibrium energy distribution that departs
   from the MB one~\cite{kaniadakis93,kaniadakis97}. 
   One should consider that constant diffusion and friction
   coefficients are only the first terms of a derivative expansion of the
   linearized hydrodynamic equations: Few percent contributions from
   the next terms in the expansion are not unrealistic, and are sufficient
   to give appreciable deviations from the MB distribution. 
   Moreover, successive scattering events should not be independent
   (Markovian) in the solar interior, but correlated over periods
   corresponding to a few scattering processes. Indeed, such correlations
   have also been found to be related to anomalous 
   diffusion~\cite{muralidhar90,wang92} and to the Tsallis' statistics that
   is a generalization of the Boltzmann-Gibbs
   statistics~\cite{tsallis88,curado91,tsallis95c}.
   Time correlations, velocity dependence of the transport coefficients or
   other deviations from the so-called Navier-Stokes limit of the hydrodynamic
   equations are all manifestations of multiple particle collisions and of
   the many-body nature of the system, which is not completely described
   by two-body short-range interactions.
   In addition, the new data of the solar neutrino experiments
   (Chlorine~\cite{lande96}, Kamiokande~\cite{fukuda96,kataup97},
   Gallex~\cite{GALLEX95,kirsten96,gataup97} and SAGE~\cite{gavrin96})
   have made the actual SNP more puzzling~\cite{bahcall97b,bahcall97a}
   and the possibility
   that Clayton's suggestion could contribute to its solution should be
   reassessed.

   It should be stressed from the beginning that modifications of the shape 
   of the energy distribution {\em are not} equivalent for solar models to
   changes of temperature and/or densities; rather, a new degree of freedom is
   introduced. For instance, one might think that the effects of a distribution
   with a depleted high-energy tail could be reproduced by a MB
   distribution with a lower temperature. However, a lower temperature
   produces two effects: (1) the rates are reduced by changing the thermal
   average $\langle v\sigma\rangle$ (most of the contribution to this average
   comes from the high-energy tail because of the particular energy
   dependence of the main cross sections $\sigma$); (2) the system finds
   a new hydrostatic equilibrium because the average momentum of the
   particles becomes smaller (particles of all energies give
   comparable contributions to this average). In contrast, a change of
   shape of the distribution can reduce the rates and, at the same time,
   maintain the same hydrostatic equilibrium, since the two effects are
  dominated by particles from different parts of the energy spectrum.
  Indeed, the effects on solar models of changing the shape of
  the energy distribution could be reproduced by simultaneous {\em local}
  changes of temperature and cross sections.

  Given the widespread misconception about the inevitability of the
 Maxwell-Boltzmann velocity distribution in the solar core, we briefly
 review in Sect.~\ref{MB} the general assumptions under which the
 MB distribution is derived, show that these assumptions can be
 only approximately verified in the solar interior, and discuss some of
 the expected corrections to the standard treatment.
 In particular we shall consider (Sect.~\ref{anomal}) three
 concrete examples of anomalous diffusion (velocity-dependent corrections
   to the diffusion coefficient, to the friction coefficient, and
   slowly-decaying velocity autocorrelation), and how they generate
   deviations from the MB energy distribution. Then in
   Sect.~\ref{solarnu} we illustrate the consequences of small deviations for
   solar neutrino fluxes and estimate the magnitude of the deviations necessary
   to change the fluxes by amounts relevant for the SNP. We reserve
   Sect.~\ref{concl} to our conclusions.

\section{The Maxwell-Boltzmann distribution and the solar core}
\label{MB}
In this Section we briefly review the standard hypotheses that lead to
the MB distribution for the single-particle velocity, and
the ones under which the two-particle relative-velocity distribution
is also Maxwellian. Then we estimate the order of
magnitude of the relevant physical quantities and show that these
hypotheses are not met, making it likely that the MB distribution be only
an approximation to the real distribution. These same ``order of magnitude''
considerations suggest some of the corrections to the hydrodynamical
equations.

\subsection{The ubiquity of the Maxwell-Boltzmann distribution}
The velocity distribution can be studied either with a dynamical kinetical
approach or with the methods of equilibrium statistical
mechanics~\cite{chapman39,landauV59,huang63,balescu91,reichl80}.

The kinetical approach yields a hierarchy of differential equations (the
Bogoliubov-Born-Green-Kirkwood-Yvon hierarchy). The $n$-th differential
equation is an equation of motion for the $n$-body distribution function
$f_n(r_1,v_1; \ldots ; r_n,v_n;
t)$, consisting of a streaming term, which involves $f_n$, and a collision
term, which involves the $(n+1)$-body distribution function. In particular,
the one-body distribution function $f_1(r_1,v_1; t)$ verifies a
differential equation that involves the two-body distribution function
$f_2(r_1,v_1; r_2,v_2; t)$ in the collision integral. The derivation of
the BBGKY hierarchy 
already assumes that {\em (1) the collision time is much smaller than the
mean time
between collisions} and that {\em (2) the interaction is sufficiently local}.
When these conditions are not met, collisions are not well-defined events in
space and time. The rate of change of the $f$'s at time $t$ would depend
not only on the $f$'s themselves at time $t$, but also on its previous
history (the process is not Markovian), and higher-order spatial
derivative cannot be discharged. Assuming that the BBGKY hierarchy is valid,
the additional hypothesis of molecular chaos (Boltzmann's Stosszahlansatz),
{\em i.e.} that {\em (3) the velocities of two particles at the same point
are not correlated} $f_2(r,v_1;r,v_2;t)= f_1(r,v_1;t)\times f_1(r,v_2;t)$,
allows the truncation of the hierarchy of equations and yields the Boltzmann
transport equation for the one-body distribution function $f_1(r,v;t)$ alone.
From the Boltzmann transport equation the Boltzmann's H theorem can be derived
whose consequences are that (i) under arbitrary initial conditions
$\lim_{t\to\infty} f(v,t) = f_0(v)$~\footnote{For the sake of this discussion,
we drop from now on the spatial label $r$, that should be reintroduced
when discussing non-local interactions or density gradients}, and that
(ii) $f_0$ is the equilibrium distribution if and only if
$f_0(v_1)f_0(v_2) = f_0(v'_1)f_0(v'_2) $, where $(v_1,v_2)$ and $(v'_1,v'_2)$
are the velocities before and after the collision, {\em i.e.}
$\sum_i\log f_0(v_i)$ is conserved in the collision. Then the additional
assumption that {\em (4) energy is locally conserved when using only the
degrees of freedom of the colliding particles} yields that the energy of
the particle contributes linearly to $\log{f_0(v)}$, and, therefore, the
one-particle equilibrium distribution is Maxwellian.

The equilibrium statistical mechanics approach uses the concept of most
probable value of the distribution, and the large number of degrees of
freedom assures that large fluctuations away from the most probable
distribution are extremely unlikely in non-critical conditions.
Probabilities are calculated using
the Postulate of Equal a Priori Probability and, when deriving
the MB distribution, the assumption {\em (3)} that the velocity probabilities
of different particles are independent. In addition, it is necessary that the
total energy of the system could be expressed as a sum of a term quadratic
in the momentum of the particle and independent of the other variables,
and a term independent of momentum~\cite{landauV59}
(this second term includes the energy of the rest of
the system and the interaction energy). This second condition is related
not only to assumption {\em (4)} but also to {\em (1)} and {\em (2)}, since
if {\em (1)} and {\em (2)} are not verified the resulting effective two-body
interaction is not local and depends on the momentum and energy of
the particles.

Finally, even when the one-particle distribution is Maxwellian, 
assumptions {\em (3)} and {\em (4)}, {\em i.e.} that the velocities are
uncorrelated and energy is locally conserved, are again necessary to deduce
that the relative-velocity distribution is also Maxwellian.

At last, we suggest two of the reasons of the ubiquity of 
the MB distribution. On the one side, whenever the above assumptions
are a good approximation (typical examples are those systems that are dilute
in the appropriate variables), the resulting equilibrium distribution can
be demonstrated to be Maxwellian independently of the details of the
interaction.
On the other side, even when conditions are such that one can not rigorously
deduce the form of the distribution, considering that most of the
experimental measurable observables do not really test the form of the
distribution, but only a few of its moments, the assumption of a MB
form has several advantages:
(a) It becomes exact in the dilute limit;
(b) Being determined by a single parameter or scale, the second moment of
the velocity distribution, it is consistent with our maximal ignorance
(it assumes the least about the distribution);
(c) It allows simple analytical treatments.

\subsection{The solar core quasi-plasma}
In the light of the above considerations, we examine the situation in the
solar core, and discuss what kind of corrections to the simplified
picture of a dilute gas  could better describe a system where, as we shall
argue, multiple particle collisions and many-body correlations between
clusters of tens of particles should be present. Of course, our discussion
suggests only qualitative answers, since these issues need a more sophisticated
theoretical framework, better microscopical models and, perhaps, they could
still be quantitatively settled only by a numerical dynamical microscopic
calculation, which is not trivial and, to our knowledge, still missing.

Ions in the solar core are completely ionized: $T_c = 1.36$~keV.

Only the many-body correlations between ions and electrons cut off the
range of the electromagnetic interaction. The Debye-H\"uckel estimate 
(weak mean-field static approach) gives a screening length 
($R_D^2=kT/(4\pi e^2 \sum_i Z^2_i n_i)$) of the order of the average
interparticle distance in the center of the Sun:
$R_D\approx n^{-1/3}\approx 3\times 10^{-9}$cm,
where $n$ is the average density.
In addition, the same Debye-H\"uckel approach estimates that, at the
average interparticle distance, {\em i.e.}, 
when the potential energy is at its
minimum, the potential energy is already about 4\% of the kinetic energy:
its contribution is comparable to the kinetic energy for large portions
of the particle trajectories.

  These ``back of the envelope'' calculations already suggest that it is not
possible to completely describe the system as streaming particles plus
collisions. Since the collision time is of the same order of the
mean time between collisions and the size of the quasi-particle (ion plus
screening cloud) is of the order of the distance between particles,
a description in terms of a gas of compact quasi-particles interacting via
local two-body collisions can only be an approximation whose corrections
involve higher order derivative terms.
In fact, the relevant expansion parameter in the BBGKY hierarchy is
$n r_0^3$, where $n$ is the density and $r_0$ the range of the interaction;
but $n r_0^3 \approx 1$ using $r_0 \sim R_D$.

Moreover, while the Debye-H\"uckel screening strictly applies only to
a charge at rest or travelling at constant velocity, {\em i.e.} in the
limit that the plasma reaction time is infinitely fast compared to the
particle rate of velocity change, the real screening is dynamical and
the inverse plasma frequency gives us an estimate of the reaction time:
$t_{pl}=\omega_{pl}^{-1} = \sqrt{m/ (4\pi n e^2 )} \approx 10^{-17}$~sec.
Times of this order of magnitude are necessary to build up screening
after a hard collision and should be compared to the typical collision
time $t_{coll} = \langle\sigma v n \rangle^{-1} \approx 10^{-17}$~sec.

   We describe the fact that $t_{pl}\approx t_{coll}$ by saying
that collective effects have a time scale comparable to the average
time between collisions and, therefore, several collisions are
necessary before the particle looses memory of the initial state.
Another consequence of $t_{pl}\approx t_{coll}$ is that collisions
between the quasi-particles (bare particle plus the screening cloud) are
inelastic: part of the energy is dissipated by the process of the creation and
successive removal of the screening cloud.

Since the plasma parameter, {\em i.e.}, the 
number of particles in the Debye sphere, is not large
($N_D\equiv(4\pi/3)R_D^3  n \approx 1$), a description in terms of a
high-temperature plasma ($1/N_D \sim n^{(1/2)} T^{-(3/2)} \to 0 $) is
not a good approximation either.
We could name it a quasi-plasma: strong many-body effects on the
scale of a few average interparticle distances appear to be necessary,
and clusters of a few (or tens) of correlated particle participate in the
collisions, the number depending on the time scale considered.

In the next Section, we show that when the lowest-order kinetic equations,
which are valid in the low-density limit, are supplemented with terms
that have the form suggested by the above
considerations, {\em i.e.} higher derivative terms (spatial nonlocality),
time correlations (temporal nonlocality), or nonlinearity, the resulting
equilibrium distribution is not Maxwellian.

\section{Anomalous diffusion, time correlations and nonstandard statistics}
\label{anomal}

For concreteness we consider three possible corrections: i) A correction
to the lowest order friction coefficient $J(v)$; ii) a correction to the lowest
order diffusion coefficient $D(v)$; and iii) a modification of the two-body
time correlations. These three types of corrections affect the
Maxwell-Boltzmann statistical distribution. Indeed, similar
corrections have already been shown to exist in hydrodynamic 
systems~\cite{dorfman72,reichl80,chaichin95}: direct microscopic
calculations could prove this possibility also in the solar context.

  It is clear that such corrections imply other consequences. For instance,
it is well-known that the present approach to the slow diffusion of heavy
elements could also be modified~\cite{paquette86,cox89}.
In the present paper, we are interested
 in only one of these consequences: the actual {\em equilibrium} statistical
 distribution of the relative-energy departs from the Maxwell-Boltzmann
 equilibrium distribution.
  
 Corrections i) and ii) have already been considered in a more general
   context~\cite{kaniadakis93,kaniadakis97}, and we only recall the main
   points. {\em We assume that the system is not too far from the
  standard regime that leads to the MB distribution, so that an expansion
starting from the usual formalism makes sense}. The Fokker-Planck equation,
given in the Landau form, is
\begin{equation}
\frac{\partial}{\partial t}f(t,v) = \frac{\partial}{\partial v}
\left( 
J(v)f(t,v) + \frac{\partial}{\partial v} D(v) f(t,v)
\right)\, ,
\end{equation}
where $f(t,v)$ is the distribution probability of particles with velocity
$v$ at time $t$ and $J(v)$ and $D(v)$ are
the dynamical friction and diffusion coefficients. 
The stationary distributions are the asymptotic solutions of the above equation.
To lowest order $J(v)= v/\tau$ and $D(v)=\epsilon/\tau$, where the constant
$\tau>0$ has dimension of time ($m / \tau$ is the friction constant) and
$\sqrt{\epsilon}$ has dimension of a velocity
($\epsilon = kT/m$ for Brownian motion).
At equilibrium one obtains the well-known Maxwellian distribution
\begin{equation}
f(v)\equiv \lim_{t\to\infty} f(t,v) \sim 
\exp\left\{-\frac{v^2}{2\epsilon}\right\} 
= \exp\left\{-\frac{m v^2}{2 kT}\right\} \, .
\end{equation}

We can generalize the standard Brownian kinetics considering the expressions
of the quantities $J(v)$ and $D(v)$ to the next order in the velocity
variable: $J(v)= v/\tau\, (1+\beta_1 v^2/\epsilon) $ and
$D(v)=\epsilon/\tau \, (1 + \gamma_1 v^2/\epsilon)$; these higher derivative
terms can be interpreted as signals of nonlocality in the Fokker-Planck
equation.

If $\beta_1= 0$ and $\gamma_1\neq0$ we find the Tsallis' distribution
\begin{equation}
\label{tsdis}
f(v) = \left[ 1 + (q-1)\frac{m v^2}{2 kT}
      \right]^{1/(1-q)}
       \Theta\left[1 + (q-1)\frac{m v^2}{2 kT}\right]\, ,
\end{equation}
where $q-1=2\gamma_1/(2\gamma_1+1)$, $\Theta$ is the
Heaviside step-function, and $kT/m \equiv \epsilon (2-q)$.
When the characteristic parameter $q$
is smaller that 1 ($-1/2<\gamma_1<0$), this distribution has a upper cut-off:
$m v^2/2 \leq kT/(1-q)$ (the tail is depleted). The distribution correctly
reduces to the exponential Maxwell-Boltzmann distribution in the limit
$q\to 1$ ($\gamma_1\to 0$).
When the parameter $q$ is greater than 1 ($\gamma_1>0$), there is no cut-off
and the (power-law) decay is slower than exponential
(the tail is enhanced).

If $\beta_1\neq 0$ and $\gamma_1=0$, we find a
Druyvenstein-like distribution:
\begin{equation}
f(v)\sim
\exp\left\{  - \frac{v^2}{2\epsilon}
            - \beta_1 \left( \frac{v^2}{2\epsilon} \right)^2 
   \right\} \, ,
\end{equation}
which has also the functional form suggested by Clayton to parameterize
a small deviation (depletion) from the Maxwellian statistics.

The statistical distribution of Eq.~(\ref{tsdis}) has an additional
appealing feature: it naturally appears in the context of the
generalized Boltzmann-Gibbs statistics obtained by introducing a new
non-extensive entropy (Tsallis' entropy):
\begin{equation}
S_q = \frac{k}{q-1}\sum_i p_i (1-p_i^{(q-1)})\, .
\end{equation}
The formal structure of the conventional thermostatistics is maintained and
its results are also naturally
generalized~\cite{tsallis88,tsallis95c,tsallis95a}. Apart from the formal
aspect,
this distribution is also attractive because of the many systems where it
plays a role~\cite{plastino93,tsallis95b,zanette95,tsallis95c,%
tsallis96b,bogho96,hamity96,rajagopal96,costa97,torres97,lavagno97}.
Non-extensivity $(q\neq 1)$ arises in systems with long-range interactions
(gravitational systems~\cite{lavagno97}, plasmas~\cite{huang94,liu94},
condensed matter~\cite{koponen97}) or with long memory
at the microscopic level.

  We mention also the possible connection between time 
  correlations and extended statistics (other mechanisms are also possible). 
  For a Markovian scattering process, the time correlation between particle
  velocities is by definition proportional to a delta function in time:
  $\langle v(0)v(t)\rangle\sim \delta(t)$.
  As discussed above, we expect that particles loose memory
  of the initial state only after a few scattering processes; we can
  model the long-time asymptotical behavior of the velocity-correlation as
  $\langle v(0)v(t)\rangle\sim t^{-(1+\gamma)}$. If $\gamma\geq 1$, {\em i.e.},
  the correlation decays sufficiently fast, the diffusion process is
  no qualitative different from the delta-function case:
  $\langle x^2(t)\rangle\sim t$. The same standard result holds if
  $0<\gamma<1$ and $0<\int\langle v(0)v(t)\rangle<\infty$. However,
  it has been shown~\cite{muralidhar90,wang92} that,
  if $0<\gamma<1$ and $\int\langle v(0)v(t)\rangle=0$ (or very small),
  or if $-1<\gamma\leq 0$, the
  diffusion is anomalous $\langle x^2(t)\rangle\sim t^{1+\gamma}$
  ($\sim t\log t$, if $\gamma=0$).
     Indeed, Tsallis~\cite{tsallis95c} shows that the generalized entropy
  $S_q$ quite naturally generate anomalous diffusion
  ($\langle x^2(t)\rangle\sim t^{1+\gamma}$): this same generalized
  entropy leads also to the non-Maxwellian probability distribution for the
  velocities given by Eq.~(\ref{tsdis}).

\section{Nonstandard statistics and solar neutrinos}
\label{solarnu}
From the considerations above we infer that generalized
distributions, among which  Tsallis' distribution has a special
theoretical appeal, could better approximate the situation in the solar
interior.

One could proceed and study the effect of using generalized distributions
on solar models, distributions with both depleted and enhanced tails,
for small and large deviations from the Maxwellian statistics.
In particular, it would be interesting to study how the relative balance
among the different reaction chains would change because of non-Maxwellian
statistics; for instance, a distribution with an enhanced high-energy
tail could make the CNO cycle important at relatively lower temperatures.
This study would be very interesting, since it could {\em experimentally}
constrain deviations from the MB statistics, given the high sensibility of
these reaction rates to the tail of the distribution.
However, such a systematic study, which should be performed by consistently
  including nonstandard statistics in solar model calculations, is not the
  purpose of this paper. 

We shall only consider small deviations from the Maxwellian distribution
and, for the purpose of illustration, use Clayton's parameterization with
the factor $e^{-\delta (E/kT)^2}$. 
 For instance, the Tsallis' distribution can be also approximated to first
order in $(1-q)$ by Clayton's form with $\delta=(1-q)/2$ and a renormalized
temperature $T'= T + T(1-q)$. The usual asymptotic expansion~\cite{clayton68}
of the integrand over the velocity distribution around the most effective
energy, $E_0$, yields an analytical expression for the rate changes.
This analytical expression is valid also for $\delta<0$ as an
asymptotic expansion around the $\delta=0$ case, in spite of the fact
that the distribution is unbounded at high energy. Instead, a numerical
integration should be performed with a suitable cut-off. In fact, the
integrand decays exponentially after the Gamow peak when $\delta\geq 0$,
and a sufficiently large cut-off does not changes the numerical value of
the integral. When $\delta > 0$, the integral still decays after the 
Gamow peak, but it goes back up at energies $E\approx E_0/\delta$: this
contribution to the enhanced tail is ``unphysical'' and comes from the
choice of the parameterization. If $\delta\ll 1$, the large window between
$E_0$ and $E_0/\delta$ allows the unambiguous elimination of this contribution.
The alternative is to use a distribution that has an enhanced tail but still
decays at high energy, such as the Tsallis' distribution for $q>1$. For
small deviations from the MB distribution, the two descriptions give the same
numerical results with the appropriate reparameterization, since one single
number characterizes the deviation to first order.

If one computes the thermal average $\langle v\sigma\rangle$ with the
modified distribution for a two-body reaction
with Coulomb barrier, one finds that to the leading order in $\delta$
\begin{equation}
\frac{\langle v\sigma_i\rangle_{\delta}}{\langle v\sigma_i\rangle_{0}}=
e^{-\delta (E_0^{(i)}/kT)^2}\equiv e^{-\delta\gamma_i}       \, ,
\label{edg}
\end{equation}
where $E_0$ is the most effective energy (maximum at the Gamow
peak)~\cite{clayton68}
\begin{equation}
\frac{E_0}{kT}\approx 5.64 \left(Z_1^2Z_2^2\frac{A_1A_2}{A_1+A_2}
            \frac{T_c}{T}\right)^{1/3} \, ,
\end{equation}
which depends on the reaction $i$ through the charges $Z$ and weights $A$ of
the ions, and on the relevant average temperature $T$. Here, $T_c=1.36$~keV is
the temperature at the center of the Sun.
In Table~\ref{table1} we report the values of $\gamma_i\equiv(E_0^{(i)}/kT)^2$
for the five most relevant reactions in the Sun:
$p+p$ ($i=1.1$), $p+^7$Be ($i=1.7$), $p+^{14}$N ($i=1.14$),
$^3$He + $^3$He ($i=3.3$) and $^3$He + $^4$He ($i=3.4$).
Changing
$\langle v\sigma\rangle$ for the $i$th reaction will affect the whole
solar model and, in general, all fluxes will change. We estimate
the effect on the fluxes by using power-law dependences
\begin{equation}
\label{dflux}
R_j\equiv\frac{\Phi_j}{\Phi_j^{(0)}}=\prod_i \left(
 \frac{\langle v \sigma_i\rangle_{\delta}}
      {\langle v \sigma_i\rangle_{0}}
                                 \right)^{\alpha_{ji}} 
     = e^{-\sum_i \delta_i\gamma_i \alpha_{ji} }  
\, ,
\end{equation}
for the fluxes $j=$ $^7$Be, $^8$B, $^{13}$N and $^{15}$O, while we have
used the solar luminosity constraint~\cite{castell97}
to determine the $pp$ flux,
$R_{pp} = 1+0.087\times(1-R_{\text{Be}})
                       +0.010\times(1-R_{\text{N}})
                       +0.009\times(1-R_{\text{O}})$, and kept fixed
the ratio $\xi\equiv\Phi_{pep}/\Phi_{pp}=2.36\times 10^{-3}$.
The exponents $\alpha_{ij}=\partial\ln\Phi_j 
  / \partial\ln\langle v \sigma_i\rangle$ (see Table~\ref{table2}) have
been taken from Ref.~\cite{castell97}, where it is
also discussed why solar models depend on
$\langle v \sigma \rangle_{33}$ and $\langle v \sigma \rangle_{34}$
only through the combination
$\langle v \sigma \rangle_{34} / \sqrt{\langle v \sigma \rangle_{33}}$
and why it is a good approximation to keep the ratio $\xi$ constant.

In principle $\delta$ should be determined by a direct calculation of the
complex many-body system and could be different for every
reaction ($\delta\to \delta_i$). The
energy distribution can be influenced by the specific properties of
the ion (charge and mass) and by the different conditions of the
environment in those parts of the Sun where each of the reactions
mostly takes place. However, a direct calculation is not simple and it does
not exist for the solar interior. Therefore, for the only purpose of
estimating
the potential effect of nonstandard distributions, we consider two simple
models and use the corresponding $\delta$('s) as free parameter(s).
The first model assumes the same deviation $\delta$ for all distributions,
while the second model assumes that only the $p+{}^7$Be relative-energy
distribution (parameterized by $\delta_{\text{Be}}$) and the two helium
reactions (parameterized by $\delta_{\text{He}}$) are non standard.

In the first case, one finds by substituting Eq.~(\ref{edg}) in 
Eq.~(\ref{dflux}) that
\begin{equation}
\frac{\Phi_j}{\Phi_j^{(0)}}=e^{-\delta \beta_j} \, ,
\end{equation}
where $\beta_j=\sum_i \alpha_{ji}\gamma_i$ are reported
in Table~\ref{table2}.
This dependence of the fluxes on $\delta$ is in good agreement with
 Clayton's numerical calculation~\cite{clayton75}.
Using the model of Ref.~\cite{bahcall95a} as reference model and the
latest experimental results (see Table~\ref{table3}), we obtain
the best fit for $\delta=0.005$ with a $\chi^2=35$.

In the second case, we proceed similarly, but we use
$\delta_{\text{Be}}$ for the reaction $p+ {}^7$Be and $\delta_{\text{He}}$
for the two reactions He + He:
the corresponding $\beta_j^{\text{Be}}=
\alpha_{ji}\gamma_i |_{i=1.7}$ and $\beta_j^{\text{He}}=
\alpha_{j,3.4}(\gamma_{3.4}-\gamma_{3.3}/2)$ are also reported in
Table~\ref{table2}. As shown in Table~\ref{table3} the best fit is
obtained for $\delta_{\text{Be}}=-0.018$ (negative $\delta$ corresponds to
an enhanced tail, $q>1$ in Tsallis' distribution)
and $\delta_{\text{He}}=0.030$ with a $\chi^2=20$.

    From previous analyses we already knew that it is not possible
   to obtain a very good fit to all experiments even when the fluxes
   are used as free
parameters~\cite{castell94b,castell94a,deglinn95,heeger96,castell97}:
    a fit that has a low probability could only be obtained by reducing
    the boron
    flux by a factor of two and the beryllium and CNO fluxes as much
    as possible. Therefore, we are not surprised that we have not been
    able to obtain good fits, however we have been able to give
   a specific and physically motivated mechanism that greatly reduces
  the discrepancy between theory and experiment (the SSM has a $\chi^2=74$).
   Another physical mechanism that produces similar results is by
   introducing arbitrary screening factors~\cite{castell97}.
   It is also consistent that the second case, which produces the best
   fit, results in a depletion of the  energy tail
    of the $^3$He and $^4$He ions, so that $^7$Be and $^8$B
    fluxes are strongly suppressed, and that
     the energy tail of the $p+^7$Be reaction is {\em enhanced} so to
   bring up the $^8$B flux towards the measured value.

    We do not claim that this result is a solution to the SNP, in the
    sense of providing a model to fit the experimental results within
    one ({\em a few}) sigma. Our point is only that deviations from
    standard statistics corresponding
    to values of $\delta$ of about 1\% can change the neutrino fluxes
    of factors comparable to those that constitute the SNP.
    Such values of $\delta$, or even larger values, cannot be excluded by
    the present knowledge of the strong interacting quasi-plasma in the
    solar interior. It could turn out that the actual values of the
    neutrino fluxes coming out of the Sun could result from the interplay
    of several mechanisms that are disregarded in the standard
    picture~\cite{dar97}.

    In the light of the above considerations, the uncertainties of the
    neutrino fluxes are considerably underestimated by not considering the
    possibility of non-extensive distributions.

    Finally, we wish to comment on the fact that limits on the reaction
    rates that come from determinations of the sound speed through
    helioseismological measurements do not automatically apply to changes
    of the rates through the present mechanism. In fact, given the cross
    sections, the reaction rates change because the densities and/or the
    thermal averages, $\langle\sigma v\rangle$, change. However, if the
    statistics is not changed, the thermal averages change when the
    temperature changes, and changes of temperature and/or density
    clearly affect the structure of the solar model and the sound speed.
    In contrast, nonstandard energy distributions make it possible to
    change the thermal averages (at least for those reactions whose main
    contribution comes from the high-energy tail) without affecting the 
    properties that depend on the bulk of the distribution, such as
    the sound speed and/or the equation of state. Therefore, non standard
    velocity
    distributions affect the helioseismological measurements only
    insomuch as the consequent changes of the rates modify the solar
    structure.

\section{Conclusions}
\label{concl}
  New developments in generalized statistics
  combined with several qualitative hints that the standard approach to
  the solar interior is only a first approximation to the
  real situation, make it worthwhile to reconsider the early suggestion
  by Clayton that the energy distribution in the Sun could depart from
  the Maxwell distribution.

  In particular, we recall that:

 (1) The conditions in the solar core (density and temperature) do not
 satisfy those requirements that would guarantee standard diffusion and
 Maxwell-Boltzmann velocity distribution.

 (2) Nonstandard diffusion is most likely present: Lowest order dynamical
 friction and diffusion coefficients are not sufficient; the diffusion
 mechanism is not described by a Markovian chain of independent
 two-body scattering events and correlations persist for time intervals
 longer than the mean time of one scattering process (memory effect).

 (3) The direct consequence of these corrections to standard diffusion
 is that the {\em equilibrium} energy distributions of electrons and ions
 are not Maxwellian. In particular, the tails of the distributions are not
 exponential and, therefore, the small number of particles that
 have energies large enough to participate in those reactions that are
 hindered by Coulomb barrier can be much less (more) than the one expected
 in the standard distribution.

 (4) Tsallis' statistics should also be considered.

 (5) Non-Maxwellian energy distributions modify the reaction rates. This
 phenomenon has the potential of increasing the range of possible values of
 the reaction rates well beyond the ones allowed by the uncertainties in the
 corresponding cross sections.

 (6) Unlike a change in the temperature, which has a direct effect on the
 hydrostatic equilibrium and on the the sound speed, modifications of the
 distribution that affect only the high-energy tail do not change the
 solar model and the sound speed: the range of neutrino fluxes from
 models that verify helioseismological constraints could also be increased.

 (7) If one modifies the standard distribution by a Clayton's factor
 $e^{-\delta (E/kT)^2}$ with $\delta$ of the order of 1\% (such modification
 cannot be excluded by the present knowledge of the strong interacting
 quasi-plasma in the solar interior) the
 neutrino fluxes change of amounts comparable to those that constitute
 the solar neutrino problem, even if it is not possible to solve the SNP
 by {\em only} modifying the energy distributions.

\acknowledgments
We thank G.~Fiorentini, G.~Mezzorani, B.~Ricci, and C.~Tsallis for
critical readings of the manuscript and informative discussions.

%
\begin{table}
\caption[taa]{
 Most effective energies for thermonuclear reactions and exponents
 $\gamma$ that characterize the change of the
 thermal average $\langle v \sigma \rangle$ to the leading order in
 $\delta$, when the energy distribution changes by a factor
 $\exp\{ -\delta (E/kT)^2 \}$:
 $\langle v \sigma \rangle_{\delta} = 
  \langle v \sigma \rangle_{0} \exp\{ -\delta\gamma \}$.
\label{table1}
             }
\begin{tabular}{rcdd}
\multicolumn{2}{c}{reaction} & 
         \multicolumn{1}{c}{$E_0/kT$} &
           \multicolumn{1}{c}{ $\gamma = (E_0/kT)^2$} \\
\tableline
 $\langle v \sigma \rangle_{11}$:  &  
        $p + p\to {^2\text{H}} + e^+ + \nu $  &  
                4.8  &  23. \\
 $\langle v \sigma \rangle_{17}$:  &
        $p + {^7\text{Be}} \to {^8\text{B}} + \gamma $  &
               13.8  &  190. \\

 $\langle v \sigma \rangle_{33}$:  &
        $^3\text{He}+  {^3\text{He}}\to \alpha + 2p $  &
               16.8  &  281. \\
 $\langle v \sigma \rangle_{34}$:  &
        $^3\text{He}+{} ^4\text{He}\to {^7\text{Be}} + \gamma $  &
               17.4  &  303. \\
$\langle v \sigma \rangle_{1,14}$:  &
       $p + {^{14}\text{N}} \to {^{15}\text{O}} + \gamma $  &
               20.2  &  407. \\
\end{tabular}
\end{table}
\begin{table}
\caption[tbb]{
The first four rows show $\alpha_{ij}=\partial\ln\Phi_{j}/
\partial\ln \langle v \sigma \rangle_{i}$, the
logarithmic partial derivative of neutrino fluxes with respect to the
parameter shown at the left of the row. These numbers are discussed in
Ref.~\cite{castell97}. The last three rows show $\beta_j$,
$\beta_j^{\text{Be}}$ and $\beta_j^{\text{He}}$, the logarithmic partial
derivative of the fluxes
with respect to the parameters $\delta$'s; as discussed in the text, they
are linear combinations of the $\alpha$'s weighted by the factors
$\gamma$ of Table~\ref{table1}.
\label{table2}
             }
\begin{tabular}{lccc}
 & $^7$Be & $^8$B & CNO \\
\tableline
$\langle v \sigma \rangle_{11}$ &
         -1.0   &   -2.7   &   -2.7   \\
$\langle v \sigma \rangle_{34} / \sqrt{\langle v \sigma \rangle_{33}}$ &
         +0.86   &   +0.92   &   -0.04   \\
$\langle v \sigma \rangle_{17}$ &
         0   &   1   &   0   \\
$\langle v \sigma \rangle_{1,14}$ &
         0   &   0   &   1   \\
\tableline
$\beta_j$ &
         117   &  277   &   338.5   \\
\tableline
$\beta^{\text{Be}}_j$ &
          0   &  190   &   0 \\
$\beta^{\text{He}}_j$ &
         140   &  150    &  -6.5   \\
\end{tabular}
\end{table}
\begin{table}
\caption[tcc]{
The first three columns show the predicted fluxes, and the
predicted gallium and chlorine signals in the SSM~\cite{bahcall95a} and
in the two models with nonstandard distribution described in the text.
The last column shows the present experimental results. For the three
models is also given the $\chi^2$ resulting by the comparison with the
experimental data.
\label{table3}
             }
\begin{tabular}{ccccc}
&  \multicolumn{3}{c}{Models}   &   \\
\cline{2-4}
&  SSM          & case I       & case II &  Experiment           \\
& ($\delta=0$)  & ($\delta=0.005$) 
                & ($\delta_{\text{Be}}=-0.018$,
$\delta_{\text{He}}=0.030$) &           \\
\tableline
$[10^9 \text{cm}^{-2} \text{s}^{-1}]$ &
             &     &      &  \\
$\Phi_{pp}$ &
        59.1  & 62.2  &  63.7  &  \\
$\Phi_{^7\text{Be}}$ &
         5.15  &  2.87  &  0.08  &  \\
$\Phi_{^{13}\text{N}}$ &
         0.62  & 0.11  &  0.75  &  \\
$\Phi_{^{15}\text{O}}$ &
         0.55  & 0.10  &  0.67  &  \\
$[10^6 \text{cm}^{-2} \text{s}^{-1}]$ &
             &     &      &  \\
$\Phi_{^8\text{B}}$ & 6.62 & 1.65  &  2.25  &
             $2.55\pm 0.21$~\tablenote{Weighted average of
         $2.80\pm 0.38$~\protect\cite{fukuda96}
     and $2.44\pm 0.26$~\protect\cite{kataup97}}\\
\tableline
[SNU] &
             &     &      &  \\
gallium & 137.0  & 100  &  97  &
   $75\pm 5$~\tablenote{Weighted average of
   $76\pm 8$~\protect\cite{gataup97} and
 $ 72\pm 13$~\protect\cite{gavrin96} }\\
chlorine &
         9.3  & 2.84  &  3.34  &
$2.54\pm 0.20$~\tablenote{Ref.~\protect\cite{lande96}} \\
\tableline
$\chi^2$ &
         74  & 35  &  20  &     \\
\end{tabular}
\end{table}

\begin{thebibliography}{10}

\bibitem[*]{email1}
Electronic address: kaniadakis@polito.it

\bibitem[\dag]{email2}
Electronic address: lavagno@polito.it

\bibitem[\ddag]{email3}
Electronic address: marcello.lissia@ca.infn.it

\bibitem[\S]{email4}
Electronic address: quarati@polito.it

\bibitem{clayton68}
D.~D.~Clayton,
{\em Principles of Stellar Evolutions and Nucleosynthesis}
(The University of Chicago Press, 1968).

\bibitem{bahcall89}
J.~N.~Bahcall,
{\em Neutrino Astrophysics} (Cambridge University Press, 1989).

\bibitem{bahcall95a}
J.~N.~Bahcall and M.~H.~Pinsonneault,
Rev. Mod. Phys. {\bf 67},  781 (1995).

\bibitem{bahcall95b}
J.~N.~Bahcall, R.~Davis, Jr., P.~Parker, A. Smirnov and R.~K~Ulrich (editors),
{\em Solar Neutrino: The First Thirty Years}
(Addison Wesley, 1995).

\bibitem{bahcall97b}
J~N.~Bahcall, M.~H.~Pinsonneault, S.~Basu and J.~Christensen-Dalsgaard,
Phys. Rev. Lett. {\bf 78}, 171 (1997).

\bibitem{fiorentini97}
B.~Ricci, V.~Berezinsky, S.~Degl'Innocenti, W.~A.~Dziembowski,
G.~Fiorentini, Phys. Lett. B {\bf 407}, 155 (1997).

\bibitem{bahcall90}
J.~N.~Bahcall and H.~A.~Bethe, Phys. Rev. Lett. {\bf 65}, 2233 (1990).

\bibitem{white93}
M.~White, L.~Krauss, and E.~Gates, Phys. Rev. Lett. {\bf 70}, 375 (1993).

\bibitem{castell93}
V. Castellani, S. Degl'Innocenti, and G. Fiorentini,
  Phys. Lett. B {\bf 303}, 68 (1993).

\bibitem{hata94a}
N.~Hata, S.~Bludman, and P.~Langacker,
Phys. Rev. D {\bf 49}, 3622 (1994).

\bibitem{castell94b}
V.~Castellani {\it et al.},
Phys. Lett. B {\bf 324}, 425 (1994).

\bibitem{bahcall94}
J.~N.~Bahcall, Phys. Lett. B {\bf 338}, 276 (1994).

\bibitem{parke95}
S.~Parke, Phys. Rev. Lett. {\bf 74}, 839 (1995).

\bibitem{bere96}
V.~Berezinsky, G.~Fiorentini, and M.~Lissia,
  Phys. Lett. B {\bf 185}, 365 (1996).

\bibitem{castell97}
V.~Castellani {\it et al.},
  Phys. Rep. {\bf 281}, 309 (1997).

\bibitem{fukugita91}
 M.~Fukugita, Mod. Phys. Lett. A  {\bf 6}, 645 (1991).

\bibitem{giunti92}
C.~Giunti, C.~W.~Kim, and U.~W.~Lee,
   Phys. Rev. D {\bf 46}, 3034 (1992).

\bibitem{GALLEX92}
GALLEX Collaboration, P.~Anselmann {\it et al.},
  Phys. Lett. B {\bf 285}, 390 (1992).

\bibitem{Bludman93}
  S.~A.~Bludman, N.~Hata, D.~C.~Kennedy, and P.~G.~Langaker,
   Phys. Rev. D {\bf 47}, 2220 (1993).

\bibitem{hata93}
  N.~Hata and P.~G.~Langaker, Phys. Rev. D {\bf 48}, 2937 (1993).

\bibitem{Krastev93}
  P.~I.~Krastev and S.~T.~Petcov,
   Phys. Lett. B {\bf 299}, 99 (1993).

\bibitem{bilenkii94}
S.~M.~Bilenkii and C.~Giunti,
   Phys. Lett. B {\bf 320}, 323 (1994).

\bibitem{fiore94}
G.~Fiorentini {\it et al.}, Phys. Rev. D {\bf 49}, 6298 (1994).

\bibitem{bilenkii95}
S.~M.~Bilenkii, A.~Bottino, C.~Giunti, and C.~W.~Kim,
    Phys. Lett. B {\bf 356}, 273 (1995).

\bibitem{Krastev96a}
  P.~I.~Krastev and S.~T.~Petcov, Phys. Rev. D {\bf 53}, 1665 (1996).

\bibitem{Krastev96b}
  P.~I.~Krastev and S.~T.~Petcov, Phys. Rev. D {\bf 54}, 7057 (1996).

\bibitem{Krastev97}
  P.~I.~Krastev and S.~T.~Petcov, Phys. Lett. B {\bf 395}, 69 (1997).

\bibitem{dar97}
A.~Dar and G.~Shaviv, Ap. J. {\bf 468}, 933 (1996);
A.~Dar, astro-ph/9707015.

\bibitem{Krauss93}
L.~M.~Krauss, E.~Gates, and M.~White, Phys. Lett. B {\bf 299}, 94 (1993).

\bibitem{Fogli94}
 G.~L.~Fogli and E.~Lisi, Astropart. Phys. {\bf 2}, 91 (1994).

\bibitem{hata94b}
N. Hata and P. Langacker, Phys. Rev. D {\bf 50}, 632 (1994).

\bibitem{fogli95}
G.~L.~Fogli and E.~Lisi, Astropart. Phys. {\bf 3}, 185 (1995).

\bibitem{bahcall96}
J.~N.~Bahcall and P.~I.~Krastev, Phys. Rev. D {\bf 53}, 4211 (1996).

\bibitem{kocha72}
G.~Kocharov, Ioffe Inst. Rep. 298 (1972), Leningrad (in Russian).

\bibitem{kocha74}
S.~Vasil'ev, G.~Kocharov, and A.~Levkovskii,
Izv. AN SSSR, Ser. Fiz. {\bf 38}, 1827 (1974).

\bibitem{kocha75}
S.~Vasil'ev, G.~Kocharov, and A.~Levkovskii,
Izv. AN SSSR, Ser. Fiz. {\bf 39}, 310 (1975).

\bibitem{clayton74}
D.~D.~Clayton, Nature {\bf 249}, 131 (1974).

\bibitem{clayton75}
D.~D.~Clayton, E.~Eliahu, M.~J.~Newman, and R.~J.~Talbot, Jr.,
Astrophys. J. {\bf 199}, 494 (1975).

\bibitem{davis73}
R. Davis, Jr. and J.~C.~Evans, in {\em Proc. XIII Int. Conf. Cosmic Rays}
     (Denver 1973).

\bibitem{bahcall69}
J.~N.~Bahcall, N.~A.~Bahcall, and R.~K.~Ulrich,
    Astrophys. J. {\bf 156}, 559 (1969).

\bibitem{haubold95a}
H.~Haubold and A.~M.~Mathai,
Astrophys. Space Sci. {\bf 228}, 77 (1995).

\bibitem{haubold95b}
H.~Haubold and A.~M.~Mathai,
Astrophys. Space Sci. {\bf 228}, 113 (1995).

\bibitem{kania96}
G.~Kaniadakis, A.~Lavagno, and P.~Quarati,
Phys. Lett. B {\bf 369}, 308 (1996).

\bibitem{quarati97}
P.~Quarati, A.~Carbone, G.~Gervino, G.~Kaniadakis, A.~Lavagno, and E.~Miraldi,
Nucl. Phys. A {\bf 621}, 345c (1997).

\bibitem{rosenbluth57}
M.~N.~Rosenbluth, W.~M.~MacDonald, and D.~L.~Judd,
Phys. Rev. {\bf 107}, 1 (1957).

\bibitem{macdonald57}
W.~M.~MacDonald and M.~N.~Rosenbluth,
Phys. Rev. {\bf 107}, 350 (1957).

\bibitem{krook76}
M.~Krook  and T.~T.~Wu,
Phys. Rev. Lett. {\bf 36}, 1107 (1976).

\bibitem{tsallis88}
C.~Tsallis, J. Stat. Phys. {\bf 52}, 479 (1988).

\bibitem{curado91}
E.~M.~F.~Curado and C.~Tsallis, J. Phys. A {\bf 24}, L69 (1991);
{\em ibid.} {\bf 24}, 3187(E) (1991); {\em ibid.} {\bf 25}, 1019(E) (1992).

\bibitem{plastino94}
A.~R.~Plastino and A.~Plastino, Phys. Lett. A {\bf 193}, 251 (1994). 

\bibitem{stariolo94}
D.~A.~Stariolo, Phys. Lett. A {\bf 185}, 262 (1994). 

\bibitem{tsallis95a}
C.~Tsallis, Physica A {\bf 221}, 277 (1995).

\bibitem{plastino93}
A.~R.~Plastino and  A.~Plastino, Phys. Lett. A {\bf 174}, 384 (1993).

\bibitem{tsallis95b}
C.~Tsallis, F.~C.~S\'a~Barreto, and E.~D.~Loh,
Phys. Rev. E {\bf 52}, 1447 (1995).

\bibitem{zanette95}
D.~Zanette and P.~Alemany, Phys. Rev. Lett. {\bf 75}, 366 (1995).

\bibitem{tsallis95c}
C.~Tsallis, {\em et al.}, Phys. Rev. Lett. {\bf 75}, 3589 (1995);
            {\em ibid.} {\bf 77}, 5442(E) (1996).

\bibitem{tsallis96b}
C.~Tsallis and  D.~J.~Bukman, Phys. Rev. E {\bf 54}, R2197 (1996).

\bibitem{bogho96}
B.~Boghosian, Phys. Rev. E {\bf 53}, 4754 (1996).

\bibitem{hamity96}
V.~H.~Hamity and D.~E.~Barraco, Phys. Rev. Lett. {\bf 76}, 4664 (1996). 

\bibitem{rajagopal96}
A.~K.~Rajagopal, Phys. Rev. Lett. {\bf 76}, 3469 (1996).

\bibitem{costa97}
U.~M.~S.~Costa, M.~L.~Lyra, A.~R.~Plastino, and C.~Tsallis,
Phys. Rev. E {\bf 56}, 245 (1997).

\bibitem{torres97}
D.~F.~Torres, H.~Vucetich, and A.~Plastino,
Phys. Rev. Lett. {\bf 79}, 1588 (1997).

\bibitem{lavagno97}
A.~Lavagno, G.~Kaniadakis, M.~Rego-Monteiro, P.~Quarati, and C.~Tsallis,
Astro. Lett. and Communications {\bf 35}, 449 (1998).

\bibitem{kaniadakis93}
G.~Kaniadakis and P.~Quarati, Physica A {\bf 192}, 677 (1993).

\bibitem{kaniadakis97}
G.~Kaniadakis and P.~Quarati, Physica A {\bf 237}, 299 (1997).

\bibitem{muralidhar90}
R.~Muralidhar, D.~Ramkrishna, H.~Nakanishi, and D.~Jacobs,
Physica A {\bf 167}, 539 (1990).

\bibitem{wang92}
K.~G.~Wang, Phys. Rev. A {\bf 45}, 833 (1992).

\bibitem{lande96}
K.~Lande, in {\it Proc. of Neutrino 96}, edited by K.~Enqvist,
K.~Huitu, and J.~Maalampi (World Scientific, Singapore, 1996), p.~25.

\bibitem{fukuda96}
Y. Fukuda {\it et al.} (Kamiokande Collaboration),
Phys. Rev. Lett. {\bf 77}, 1683 (1996).

\bibitem{kataup97}
K.~Inoue (Super-Kamiokande Collaboration),
preliminary result presented at Int. Workshop TAUP97,
Sept. 7--11, 1997, Laboratori Nazionali del Gran Sasso, Assergi (Italy).

\bibitem{GALLEX95}
GALLEX Collaboration, P.~Anselmann {\it et al.},
  Phys. Lett. B 342 (1995) 440; {\it ibid.} {\bf 361}, 235(E) (1996).

\bibitem{kirsten96}
 T.~Kirsten {\it et al.} (GALLEX Collaboration),
{\it Neutrino 96, Proceedings of the 17th International
Conference on Neutrino Physics and Astrophysics, Helsinki, Finland,
 13--19 June 1996}, edited by K. Huitu, K. Enqvist, and J. Maalampi 
(World Scientific, Singapore, in press).

\bibitem{gataup97}
Preliminary result presented at TAUP97,
Sept. 7--11, 1997, Laboratori Nazionali del Gran Sasso, Assergi (Italy).

\bibitem{gavrin96}
 V. Gavrin {\em et al.} (SAGE Collaboration), 
in {\em Neutrino 96, Proceedings of the 17th International
Conference on Neutrino Physics and Astrophysics, Helsinki, Finland,
 13--19 June 1996}, edited by K. Huitu, K/ Enqvist, and J. Maalampi 
(World Scientific, Singapore, in press).

\bibitem{bahcall97a}
 J.~N.~Bahcall, in {the Proceedings of
the 18th Texas Symposium on Relativistic Astrophysics, December
15--20, 1996, Chicago, Illinois}, edited by A. Olinto, J. Frieman,
and D. Schramm (World Scientific, Singapore, in press).

\bibitem{chapman39}
S.~Chapman and T.~G.~Cowling, {\em The Mathematical Theory of Non-Uniform
Gases} (Cambridge University Press, Cambridge, 1939).

\bibitem{landauV59}
L.~D.~Landau and E.~M.~Lifshitz, {\em Statistical Physics, Part 1}
(Pergamon Press, Oxford, 1980)

\bibitem{huang63}
K.~Huang, {\em Statistical Mechanics}
(John Wiley \& Sons, Inc., New York, 1963);
2nd Edition (1987).

\bibitem{balescu91}
R.~Balescu, {\em General Concepts of Statistical Mechanics}
(Krieger Publishing Company, Malabar, Florida, 1991)

\bibitem{reichl80}
L.~E.~Reichl, {\em A Modern Course in Statistical Physics}
(University of Texas Press, Austin, 1980); 2nd Edition
(John Wiley \& Sons, Inc., New York, 1998)

\bibitem{dorfman72}
J.~Dorfman and E.~Cohen, Phys. Rev.~A {\bf 6}, 776 (1972).

\bibitem{chaichin95}
P.~Chaichin and T.~Lubensky, {\em Principles of Condensed Matter Physics}
(Cambridge University Press, Cambridge, 1995).

\bibitem{paquette86}
C.~Paquette, C.~Pelletier, G.~Fontaine, and G.~Michaud,
  Ap.~J. Suppl. {\bf 61}, 177 (1986).

\bibitem{cox89}
A.~N.~Cox, J.~A.~Guzik, and R.~B.~Kidman,
Astrophys. J. {\bf 342}, 1187 (1989), and Refs. therein.

\bibitem{huang94}
X.~Huang and C.~Driscoll, Phys. Rev. Lett. {\bf 72}, 2187 (1994).

\bibitem{liu94}
J.~Liu, J.~De Groot, J.~Matte, T.~Johnston, and R.~Drake,
  Phys. Rev. Lett. {\bf 72}, 2717 (1994).

\bibitem{koponen97}
I.~Koponen, Phys. Rev. E {\bf 55}, 7759 (1997), and Refs. therein.

\bibitem{castell94a}
V.~Castellani {\em et al.}, Phys. Rev. D {\bf 50}, 4749 (1994).

\bibitem{deglinn95}
S.~Degl'Innocenti, G.~Fiorentini, and M. Lissia,
Nucl. Phys. Proc. Suppl. {\bf 43}, 66 (1995).

\bibitem{heeger96}
K.~M.~Heeger and R.~G.~H.~Robertson, Phys. Rev. Lett. {\bf 77}, 3720 (1996).

\end{thebibliography}
\end{document}